\documentclass[]{aastex62}
\usepackage{times,graphicx,amssymb,color}

\shorttitle{Orbital alignment of MBCs}
\shortauthors{Kim, JeongAhn, \& Hsieh}

\begin{document}

\title{Orbital Alignment of Main-Belt Comets}

\correspondingauthor{Yoonyoung Kim}
\email{yoonyoung@astro.snu.ac.kr}

\author{Yoonyoung Kim}
\affil{Department of Physics and Astronomy, Seoul National University, Gwanak, Seoul 08826, Korea}

\author{Youngmin JeongAhn}
\affil{Korea Astronomy and Space Science Institute, 776 Daedeokdae-ro, Yuseong-gu, Daejeon 34055, Korea}
\affil{Instituto de Astronom\'ia, Universidad Nacional Aut\'onoma de M\'exico, Apdo. Postal 106, Ensenada, B.C. 22860 M\'exico}

\author{Henry H. Hsieh}
\affil{Planetary Science Institute, 1700 East Fort Lowell Rd, Suite 106, Tucson, AZ 85719, USA}
\affil{Institute of Astronomy and Astrophysics, Academia Sinica, P.O. Box 23-141, Taipei 10617, Taiwan}

\begin{abstract}

We examine the orbital element distribution of main-belt comets (MBCs), which are objects that exhibit cometary activity yet orbit in the main asteroid belt, and may be potentially useful as tracers of ice in the inner solar system. We find that the currently known and currently active MBCs have remarkably similar longitudes of perihelion, which are also aligned with that of Jupiter. The clustered objects have significantly higher current osculating eccentricities relative to their proper eccentricities, consistent with their orbits being currently, though only temporarily, secularly excited in osculating eccentricity due to Jupiter's influence. At the moment, most MBCs seem to have current osculating elements that may be particularly favorable for the object becoming active (e.g., maybe because of higher perihelion temperatures or higher impact velocities causing an effective increase in the size of the potential triggering impactor population). At other times, other icy asteroids will have those favorable conditions and might become MBCs at those times as well.

\end{abstract}


\section{Introduction}

Active asteroids are a population of small solar system bodies that show comet-like dust emission but orbit in the main asteroid belt \citep{Jewitt2015}.
They have Tisserand parameter values of $T_\mathrm{J}$$\,>\,$3 (i.e., they are dynamically decoupled from Jupiter), distinct from the Kuiper belt comets and Oort cloud comets, which have $T_\mathrm{J}$$\,<\,$3.
Included within the active asteroid population are main-belt comets \citep[MBCs;][]{Hsieh2006}, whose activity is believed to be driven by sublimation of ice, and disrupted asteroids, whose activity is caused by other processes such as impacts or rotational destabilization \citep{Jewitt2010,Hsieh2012}.
Sublimation from MBCs suggests that ice is present today in the asteroid belt, suggesting that we may be able to use them to trace the present-day distribution of ice in the main asteroid belt. Coupled with a understanding of the dynamical evolution undergone by small bodies since the early solar system \citep[e.g.,][]{Walsh2011}, this information could then place constraints on the distribution of ice in the primordial solar system, and in turn, constrain protosolar disk models and provide insights into terrestrial water delivery \citep[e.g.,][]{Morbidelli2000,Raymond2004,OBrien2006}.

About 20 years since the discovery of the first known MBC in 1996 \citep{Elst1996}, MBCs are still rare, with about 10 objects currently to date (out of a total of $\sim$20 currently known active asteroids).
Due to the small number of known MBCs and the diversity of objects within that small sample, it has been difficult to make progress in understanding their properties as a population.
As such, finding more MBCs is a key component in efforts to improve our understanding of characteristics of the population (such as distribution of nucleus sizes, and ranges of activity strength and duration of activity) and the distribution and abundance of ice in the inner solar system \citep[cf.][]{Snodgrass2017}.

In this work, we examine the current osculating elements and proper elements of the currently known and currently active MBCs and identify trends in their distribution.
We then describe the possible significance of these trends for proposed activation mechanisms for MBCs, discuss observability and discoverability, and suggest observing strategies for discovering more MBCs based on our results.

\section{Orbital element analysis}

\subsection{Longitude of perihelion clustering}

\begin{deluxetable}{lclrrrrrrrrrrrr}
 \tablecaption{The currently known and currently active MBCs in the outer main belt (as of 2017 September).\label{tab:mbc}}
\tablehead{
\colhead{Object} & \colhead{Mech.$^{a}$} & \colhead{Disc. date$^{b}$} & \colhead{$a_{\rm osc}^{c}$} & \colhead{$e_{\rm osc}^{d}$} & \colhead{$i_{\rm osc}^{e}$} & \colhead{$\Omega_{\rm osc}^{f}$} & \colhead{$\omega_{\rm osc}^{g}$} & \colhead{$\varpi_{\rm osc}^{h}$} & \colhead{${l_\mathrm{dis}}^{i}$} & \colhead{${b_\mathrm{dis}}^{j}$} & \colhead{${a_{\rm p}}^{k}$} & \colhead{${e_{\rm p}}^{l}$} & \colhead{$\sin{i_{\rm p}^{m}}$} & \colhead{${\Delta e}^{n}$} \\
  & & & \colhead{(au)} & & \colhead{(deg)} & \colhead{(deg)} & \colhead{(deg)} & \colhead{(deg)} & \colhead{(deg)} & \colhead{(deg)} & \colhead{(au)} & &
  }
  \startdata
 133P$^{\ast}$ & S/R & 1996 Jul 14 & 3.166 & 0.159 & 1.39 & 160.1 & 131.4 & -68.5 & -44.8 & 0.6 & 3.164 & 0.153 & 0.024 & 0.006 \\
 176P & S & 2005 Nov 26 & 3.196 & 0.193 & 0.23 & 345.9 & 35.4 & 21.3 & 32.9 & 0.2 & 3.218 & 0.144 & 0.024 & 0.048 \\
 238P$^{\ast}$ & S & 2005 Oct 24 & 3.164 & 0.252 & 1.26 & 51.7 & 324.9 & 16.5 & 43.9 & -0.2 & 3.179 & 0.209 & 0.017 & 0.043 \\
 288P$^{\ast}$ & S & 2011 Nov 05 & 3.049 & 0.201 & 3.24 & 83.2 & 281.0 & 4.2 & 35.8 & -2.4 & 3.054 & 0.160 & 0.038 & 0.041 \\
 313P$^{\ast}$ & S & 2014 Sep 24 & 3.163 & 0.239 & 10.95 & 106.3 & 254.0 & 0.3 & 7.9 & -10.8 & 3.152 & 0.206 & 0.179 & 0.033 \\
 324P$^{\ast}$ & S & 2010 Sep 14 & 3.095 & 0.154 & 21.42 & 270.6 & 58.4 & -31.0 & -10.5 & 21 & 3.100 & 0.115 & 0.382 & 0.039 \\
 358P & S & 2012 Oct 06 & 3.150 & 0.237 & 11.06 & 85.7 & 299.9 & 25.6 & 34.4 & -8.7 & 3.161 & 0.196 & 0.176 & 0.041 \\
 P/2013 R3 & S/R & 2013 Sep 15 & 3.034 & 0.275 & 0.87 & 342.1 & 10.9 & -7.0 & 4.9 & 0.3 & 3.028 & 0.223 & 0.035 & 0.051 \\
 P/2016 J1 & S/R & 2016 May 05 & 3.173 & 0.228 & 14.33 & 199.9 & 46.6 & -113.5 & -128.5 & 7.6 & 3.165 & 0.260 & 0.249 & -0.031 \\
\enddata
\tablecomments{
{$^{\ast}$MBCs observed to exhibit recurrent activity. $^{a}$Likely activity mechanism (S: sublimation; S/R: sublimation and rotation).}
{$^{b}$UT date of discovery of a new MBC. $^{c}$Osculating semimajor axis. $^{d}$Osculating eccentricity. $^{e}$Osculating inclination.}
{$^{f}$Osculating longitude of ascending node. $^{g}$Osculating argument of perihelion. $^{h}$Osculating longitude of perihelion.}
{$^{i}$Ecliptic longitude at discovery. $^{j}$Ecliptic latitude at discovery.}
{$^{k}$Proper semimajor axis. $^{l}$Proper eccentricity. $^{m}$Sine of proper inclination.
$^{n}$Difference between the current osculating eccentricity and the proper eccentricity ($\Delta e = e_{\rm osc}-e_{\rm p}$).}}
\end{deluxetable}

\begin{figure}
\epsscale{0.5}
\begin{center}
\plotone{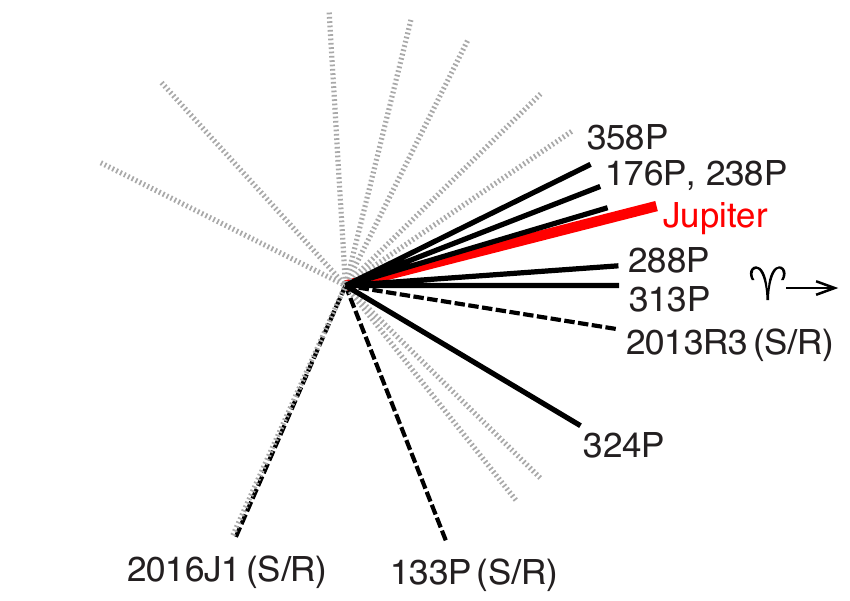}
\caption{A graphical representation of the directions of longitudes of perihelion ($\varpi_{\rm osc}$) of all known active asteroids in the main belt, where the $\varpi_{\rm osc}$ directions of all MBCs listed in Table \ref{tab:mbc} are labeled. Black solid lines correspond to sublimation-driven MBCs (S), dashed lines correspond to MBCs whose activity may be partly rotationally-driven (S/R), gray dotted lines correspond to other active asteroids (including MBCs in the middle main belt and disrupted asteroids), and a red solid line corresponds to Jupiter. The direction of the vernal equinox (ecliptic longitude of 0$\degr$) is indicated.\label{fig:orbit}}
\end{center} 
\end{figure}

A previous examination of the orbital element distribution of the known MBCs was performed by \citet{Hsieh2015},
who found an excess of high osculating eccentricities of MBCs relative to the background outer main belt asteroid population.
As an extension of their study, which mainly focused on three osculating orbital elements (semimajor axis, $a_{\rm osc}$, eccentricity, $e_{\rm osc}$, and inclination, $i_{\rm osc}$),
we explore the current osculating angular element distribution of MBCs in terms of the longitude of ascending node, $\Omega_{\rm osc}$, the argument of perihelion, $\omega_{\rm osc}$, and the longitude of perihelion, $\varpi_{\rm osc}=\Omega_{\rm osc}+\omega_{\rm osc}$.
The distributions of these angular elements have not been considered important in the past, but have recently attracted attention for their roles in implying the existence of an unknown distant planet and explaining the seasonal variation of the impact flux on Mars \citep{Trujillo2014,JeongAhn2015}.

We limit our analysis here to the outer main asteroid belt (OMB) population (defined as the population of objects with $2.824 < a_{\rm osc} \le 3.277$ au, $e_{\rm osc} \le 0.45$, and $i_{\rm osc} \le 40 \degr$), following \citet{Hsieh2015}, where nearly all known MBCs are found.
There are 493,574 multiopposition asteroids whose synthetic proper elements are available at the AstDys website ({\tt http://hamilton.dm.unipi.it/astdys/}) as of 2014 November.
We obtained osculating orbital elements of these asteroids from the Minor Planet Center's MPC Orbit (MPCORB) database at epoch 2017 September 4 and selected OMB objects from this sample using the criteria described above, resulting in a remaining sample of 114,345 OMB asteroids whose proper elements are known (hereafter background OMB population).

The currently known and currently active MBCs that are found in the OMB (as of 2017 September) are listed in Table \ref{tab:mbc}, along with their orbital elements and discovery circumstances.
The osculating elements were retrieved from the Minor Planet Center at epoch 2017 September 4. The proper elements provided by B.\ Novakovi{\'c} \citep[2016, private communication; also presented in][]{Hsieh2018} are computed using the same method used to compute the synthetic proper elements at the AstDys website using the method described by \citet{Knezevic2000,Knezevic2003}.
For our purposes, we include all objects for which sublimation has been determined to play at least a partial role in observed activity \citep[cf.][]{Jewitt2015}.
While osculating orbital elements change over time of course, they typically do not evolve significantly on timescales of $\sim$25 years, during which time all of the currently known MBCs have been seen to be active.  As such, we note that we are explicitly considering osculating elements coinciding with each object's period of observed activity.

It is interesting to note that the longitudes of perihelion of most of the currently known and currently active MBCs are clustered near $\varpi_{\rm osc}$$\,\sim\,$0$\degr$ (cf.\ Figure \ref{fig:orbit}).
While the clustering of $\varpi_{\rm osc}$ values for MBCs whose activity may be sublimation-driven is evident (black solid lines in Figure \ref{fig:orbit}; $\langle\varpi_\mathrm{MBC}\rangle$$\,\sim\,$$\varpi_\mathrm{J}$), the $\varpi_{\rm osc}$ values of MBCs whose activity may be partly rotationally-driven (i.e., 133P, P/2013 R3, and P/2016 J1) have a somewhat more dispersed angular distribution (dashed lines) and the $\varpi_{\rm osc}$ values of other active asteroids (including non-OMB MBCs and disrupted asteroids) are significantly more dispersed (gray dotted lines).

The Rayleigh $z$ test was used to quantify the significance of the longitude of perihelion clustering of the known MBC population.
By this test there is only a 0.3\% chance that $\varpi_{\rm osc}$ values of MBCs are uniformly distributed, indicating that these values are clustered at a 3$\sigma$ significance level.
The $\Omega_{\rm osc}$ and $\omega_{\rm osc}$ values of MBCs do not show statistically significant clustering with the given small sample size.
While it is well-known that $\varpi_{\rm osc}$ values for the background OMB population are also somewhat clustered near $\varpi_{\rm osc}$$\,\sim\,$$\varpi_\mathrm{J}$ \citep[e.g.,][]{JeongAhn2014}, the Watson's {\textit U}$^{2}$ test shows that the probability that the $\varpi_{\rm osc}$ distributions of the MBCs and background OMB population are drawn from the same parent distribution is $<$0.02.
Thus, the difference is statistically significant, indicating that $\varpi_{\rm osc}$ values of the MBCs are even more clustered than the those of the background OMB population.

\subsection{Secular excitation}

\begin{figure}
\epsscale{0.5}
\begin{center}
\plotone{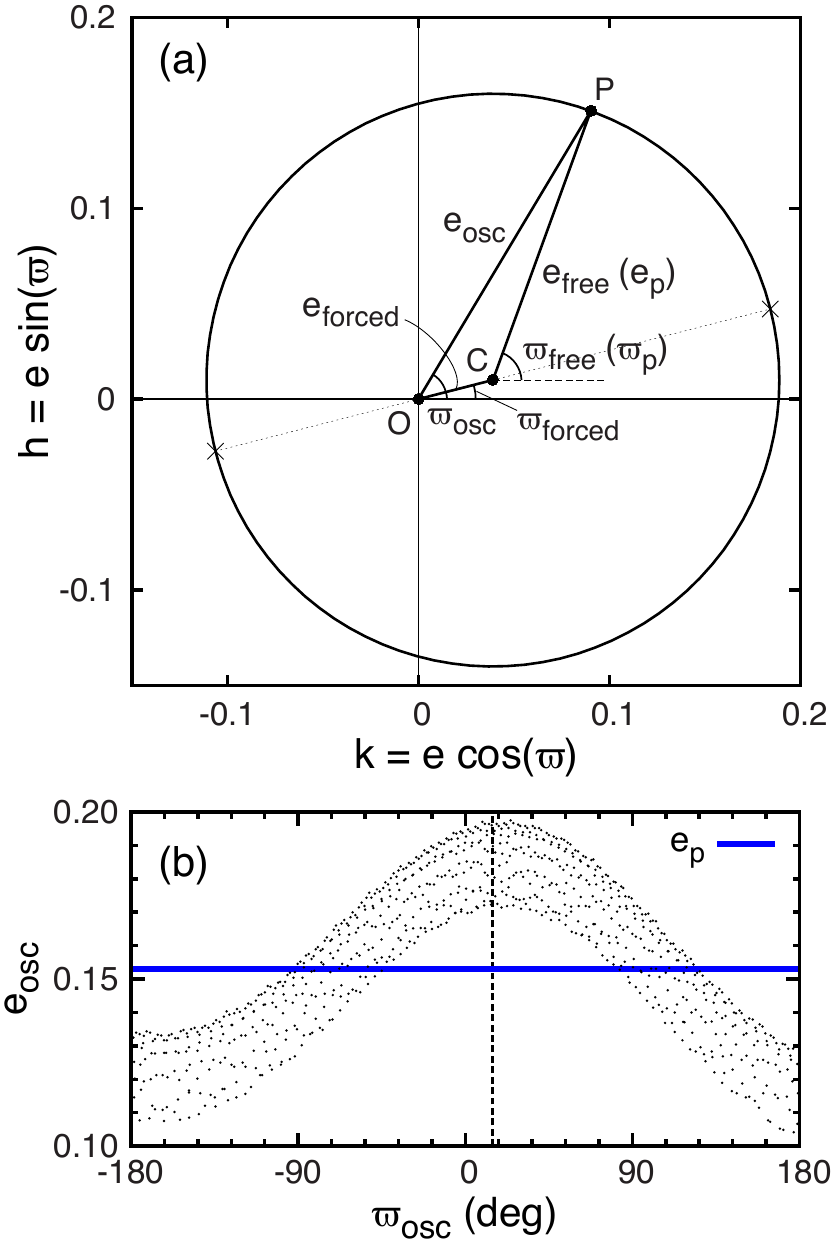}
\caption{(a) The geometrical relationship among the osculating, free, and forced eccentricities and longitudes of perihelion for a typical case of MBC ($e_{\rm p}$$\,>\,$$e_{\rm forced}$).
In this case, the ``free'' oscillation with amplitude of $e_{\rm p}$ based on the forced term (point C) results in the apparent oscillation of $e_{\rm osc}$ with amplitude of $e_{\rm forced}$ based on $e_{\rm p}$.
The maximum and minimum $e_{\rm osc}$ positions are also marked with crosses.
(b) Example of the apparent $e_{\rm osc}$ oscillation during the orbital evolution of MBC 133P over $\sim$8,000 years with output values computed at 10-year intervals.
$e_{\rm osc}$ is maximized when $\varpi_\mathrm{osc}$ is aligned with $\varpi_\mathrm{forced}$$\,\sim\,$$\varpi_{\rm J}$ (vertical dashed line).\label{fig:secular}}
\end{center} 
\end{figure}

\begin{figure}
\epsscale{1.15}
\begin{center}
\plotone{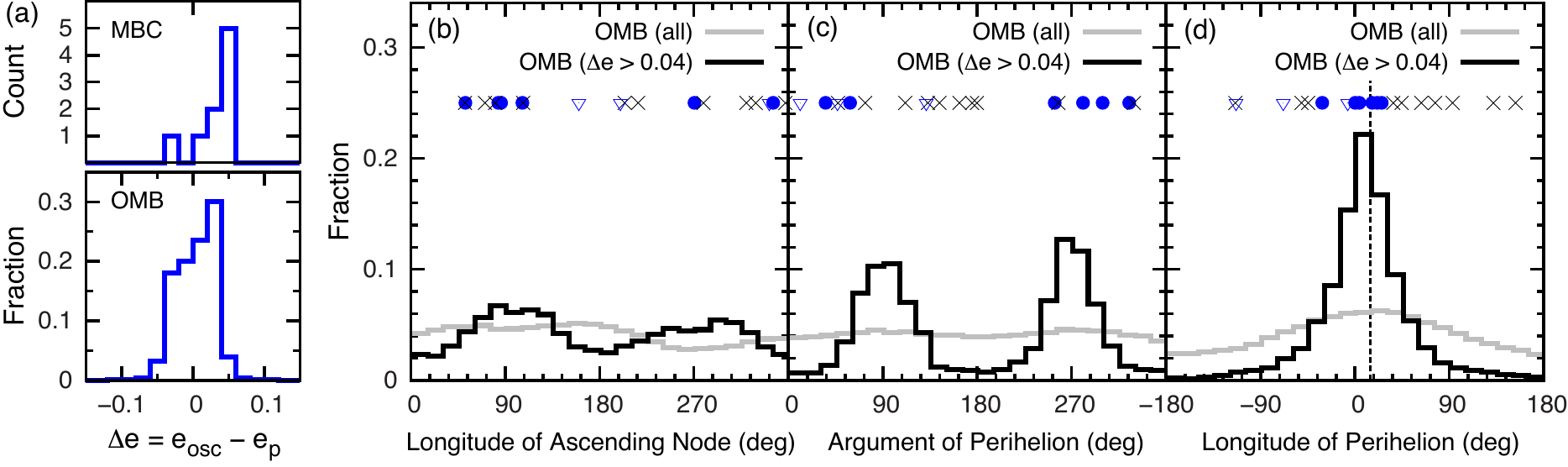}
\caption{(a) The $\Delta e$ distributions of currently active MBCs and background OMB population.
(b)--(d) The distributions of the (b) longitudes of ascending node ($\Omega_{\rm osc}$), (c) arguments of perihelion ($\omega_{\rm osc}$), and (d) longitudes of perihelion ($\varpi_{\rm osc}$) of OMB asteroids whose proper elements are known. Gray histograms denote the entire population and black histograms denote the population on osculating eccentricity-excited orbits ($\Delta e > 0.04$).
In panel (d), Jupiter's longitude of perihelion, $\varpi_\mathrm{J}\simeq15\degr$, is indicated by the dashed line.
The current angular elements of known outer main-belt MBCs (blue circles) including the MBCs whose activity may be partly rotationally-driven (blue triangles), and other active asteroids (black crosses) are also plotted.\label{fig:omb}}
\end{center} 
\end{figure}

The observed $\varpi_{\rm osc}$-clustering of the currently known and currently active OMB MBC population are consistent with being caused by secular perturbations from planets and observational selection effects. In linear secular theory, where short-period perturbations are averaged out in the disturbing function \citep{ssd}, $e_{\rm osc}$ and $i_{\rm osc}$ of a perturbed object (e.g., an asteroid) undergo secular evolution coupled with variations in $\varpi_{\rm osc}$ and $\Omega_{\rm osc}$, respectively. In the plane $(k,h)$$\,=\,$$(e_{\rm osc}\cos{\varpi_{\rm osc}}, e_{\rm osc}\sin{\varpi_{\rm osc}})$, $e_{\rm osc}$ and $\varpi_{\rm osc}$ can be obtained from the vector sums of the forced and free terms, i.e., $(k,h)$$\,=\,$$(e_\mathrm{forced} \cos{\varpi_\mathrm{forced}}+e_\mathrm{free} \cos{\varpi_\mathrm{free}}, e_\mathrm{forced} \sin{\varpi_\mathrm{forced}}+e_\mathrm{free} \sin{\varpi_\mathrm{free}})$. Values for forced terms vary over time and are determined by planetary perturbations at the given semimajor axis of the perturbed object.  Meanwhile, $e_\mathrm{free}$ is constant and $\varpi_\mathrm{free}$ uniformly precesses over time (Figure \ref{fig:secular}a). Therefore, when $\varpi_\mathrm{free}$ is aligned with $\varpi_\mathrm{forced}$, $e_{\rm osc}$ is maximized. Values for $i_{\rm osc}$ and $\Omega_{\rm osc}$ evolve in the same way as $e_{\rm osc}$ and $\varpi_{\rm osc}$ in the plane $(q, p)$$\,=\,$$(i_{\rm osc}\cos{\Omega_{\rm osc}}, i_{\rm osc}\sin{\Omega_{\rm osc}})$, meaning that $i_{\rm osc}$ is maximized when $\Omega_{\rm osc}$, $\Omega_\mathrm{forced}$, and $\Omega_\mathrm{free}$ are all aligned. We note that ``free'' elements are also referred to as ``proper'' elements and widely used in determination of asteroid families \citep[cf.][]{Nesvorny2015}. 

For OMB asteroids, Jupiter is the dominant perturbing body, and therefore $\varpi_\mathrm{forced}$ and $\Omega_\mathrm{forced}$ are directed approximately towards Jupiter's longitude of perihelion ($\varpi_\mathrm{J}$$\,\simeq\,$15$\degr$) and longitude of ascending node ($\Omega_\mathrm{J}$$\,\simeq\,$101$\degr$), respectively. 
As a result, $\varpi_{\rm osc}$ for an asteroid tends to align with $\varpi_\mathrm{J}$ when $e_{\rm osc}$ is maximized (e.g., Figure \ref{fig:secular}b), while $\Omega_{\rm osc}$ tends to align with $\Omega_\mathrm{J}$ when $i_{\rm osc}$ is maximized by the secular perturbation of Jupiter.
These theoretical conditions suggest that the orbits of the $\varpi_{\rm osc}$-aligned MBCs are currently (though only temporarily) excited in osculating eccentricity, where they have been discovered when their osculating eccentricities over the course of secular cycle are near the maximum.
For the case $e_{\rm p}$$\,>\,$$e_{\rm forced}$, the maximum and minimum $e_{\rm osc}$ during the secular cycle are $e_{\rm p}$$\,+\,$$e_{\rm forced}$ and $e_{\rm p}$$\,-\,$$e_{\rm forced}$, respectively (cf. Figure \ref{fig:secular}a). 
The maximum and minimum $\Delta e$ values (which are equivalent to $e_{\rm osc}$$\,-\,$$e_{\rm p}$) are thus $e_{\rm forced}$ and $-$$e_{\rm forced}$, respectively.
We find the forced eccentricity of $e_\mathrm{forced}$$\,\sim\,$0.04 at 3~au \citep[cf.\ Figure 7.5 in][]{ssd}.
Therefore, the $\Delta e$ value near $e_\mathrm{forced}$$\,\sim\,$0.04 can be a useful indicator of the secularly maximized osculating eccentricity for OMB objects and we can see this empirically for the MBCs (discussed below).

To quantitatively determine the degree of current secular excitation of the currently known and currently active MBCs, we calculated the differences between their current osculating elements and proper elements (Table \ref{tab:mbc}).
While generally speaking, there are negligible differences in osculating and proper semimajor axis values, we find significant differences between the current osculating eccentricities and proper eccentricities ($\Delta e$$\,\sim\,$0.04) of most of the MBCs when $\varpi_{\rm osc} \sim \varpi_\mathrm{J}$.
At 3.1~au (the approximate semimajor axis of MBCs), $\Delta e$ values of MBCs result in significant differences between the current osculating perihelion distances and proper perihelion distances ($\Delta q$$\,\sim\,$$-$0.12~au).
We also find that three of the currently active MBCs (288P, 313P, and 358P) are on orbits that are secularly excited in osculating inclination as well as in osculating eccentricity ($\Delta e$$\,>\,$0.03 and $\Delta \sin(i)$$\,>\,$0.01), but that such current secular osculating inclination excitation is not consistently shown for the currently known and currently active MBCs. Therefore, the secular osculating inclination excitation does not seem to have a significant effect on the current activity of MBCs as much as the secular osculating eccentricity excitation.

To determine the significance of the current secular osculating eccentricity excitation found for the MBCs listed in Table \ref{tab:mbc}, we compare the distribution of their $\Delta e$ values and angular elements to those of the background OMB population.
Figure \ref{fig:omb}a shows an excess of high $\Delta e$ values of MBCs relative to the background OMB population. The Kolmogorov-Smirnov (KS) test shows that the likelihood that the $\Delta e$ distributions of the currently active MBCs and background OMB population are drawn from the same parent distribution is $<$0.001 ($>$3$\sigma$ significance level), confirming that our finding that currently active MBCs tend to be secularly excited in osculating eccentricity relative to the background population is statistically significant.
As a result, $e_{\rm osc}$ values of currently active MBCs tend to be higher than those of the background OMB population as previously noted in \citet{Hsieh2015}.

Figure \ref{fig:omb}b-d shows the generally uniform distributions of current osculating angular elements for the overall OMB asteroid population (gray histograms) and the significantly non-uniform distributions of current osculating angular elements for objects on high-$\Delta e$ orbits (black histograms).
The alignment of $\varpi_{\rm osc}$ values for osculating eccentricity-excited objects around $\varpi_\mathrm{J}$ is evident (Figure \ref{fig:omb}d), in agreement with the observed $\varpi_{\rm osc}$ alignment of currently active MBCs (blue circles), while the current osculating angular elements of other active asteroids (black crosses) are more broadly distributed.
The osculating eccentricity of an object is maximized when $2\omega_{\rm osc}$$\,=\,$$180\degr$, where $\omega_{\rm osc}$ circulates due to secular perturbations \citep{Kozai1962,JeongAhn2014}, which explains the clustering of $\omega_{\rm osc}$ values at 90$\degr$ and 270$\degr$ for osculating eccentricity-excited objects (Figure \ref{fig:omb}c, black histogram).
This combination of $\omega_{\rm osc}$-clustering at 90$\degr$ and 270$\degr$ and $\varpi_{\rm osc}$-clustering near 0$\degr$ then results in a weak concentration of $\Omega_{\rm osc}$ values (which are equivalent to $\varpi_{\rm osc}-\omega_{\rm osc}$) near 90$\degr$ and 270$\degr$ (Figure \ref{fig:omb}b, black histogram).

In summary, our analysis confirms that the distribution of current osculating angular elements of OMB objects (including MBCs and background asteroids) is consistent with being largely modulated by secular perturbations from Jupiter. More interestingly, however, we find that if MBCs tend to be consistently secularly excited in osculating eccentricity (as we discuss above), they will also tend to have longitudes of perihelion close to that of Jupiter ($\varpi_{\rm osc}$$\,\approx\,$15$\degr$).

\bigskip
\section{Discussion}

\subsection{Implications of high osculating eccentricity values for MBCs}

It is hypothesized that MBCs contain subsurface ice accreted early in the solar system's history and have been ``activated'' (i.e., have had that ice exposed, triggering sublimation) by recent events such as impacts \citep[e.g.,][]{Hsieh2006,Schorghofer2008}.
If many, if not all, OMB asteroids have the potential to be MBCs (i.e., are icy), as has been suggested by thermal modeling \citep[e.g.,][]{Schorghofer2008}, then this suggests that objects with larger than average osculating eccentricities are either more likely to be currently active, are more easily discovered when active, or both.
Here we consider the possible implications of the high $e_{\rm osc}$ values for currently active MBCs (as a result of the current secular osculating eccentricity excitation) for proposed activation mechanisms and discoverability.

In the commonly considered case of collisional activation, theoretical calculations show that collision rates are nearly independent of $e_{\rm osc}$,
where it is typically assumed that the observed asteroid population (which consists of relatively large asteroids) is representative of the impactor population (consisting of much smaller asteroids) \citep{Farinella1992}.
On the other hand, higher $e_{\rm osc}$ values result in higher impact velocities, which would allow smaller impactors to still penetrate deeply enough to trigger activity \citep[cf.][]{Haghighipour2016}.
Thus, even if collision probabilities remain constant, the inclusion of smaller asteroids among the viable impactor population for asteroids on orbits with higher $e_{\rm osc}$ values means that we should expect more frequent activity-triggering collisions for such objects, and therefore higher collisional activation rates.  A quantitative analysis of this increased collisional activation rate for high-$e_{\rm osc}$ objects is beyond the scope of this present work, but will be essential for understanding the importance of this effect.

There are of course several other activation mechanisms that have been proposed for MBCs.
For example, torques from anisotropic radiation \citep[the Yarkovsky-O'Keefe-Radzievskii-Paddack effect, or YORP;][]{Rubincam2000} or anisotropic outgassing \citep[cf.][]{Belton2011} can drive a small body into rotational instability \citep[cf.][]{Jewitt2017}, and thermal stresses on the surface material may cause surface landslides or avalanches, triggering sublimation-driven activity \citep[cf.][]{Grun2016,Pajola2017}.
Unlike for collisions, it is less obvious how higher $e_{\rm osc}$ values may affect the efficacy of these mechanisms in activating MBCs. Future investigations into the relevance of the secular osculating eccentricity excitation to those mechanisms could be useful.

Turning to issues of observability and discoverability, we also consider the possible effect of higher osculating eccentricity on the production and persistence of activity for currently active MBCs after an activation event has exposed ice to solar heating.
\citet{HS15} pointed out that the currently known and currently active MBCs are rarely seen to remain observably active beyond 2.8 au (where heliocentric distance appears to be the main modulator of MBC activity).
We also note that all currently active MBCs have $q_{\rm osc}<2.7$ au.
These trends suggest that the activity of MBCs with large perihelion distances ($q_{\rm osc}\gtrsim2.7$~au) may be on the edge of detectability.

Considering the energy balance equation for a sublimating gray body at a given distance from the Sun \citep[cf.][]{Hsieh2015}, we compute the water sublimation rate at perihelion for objects with 133P's semimajor axis ($a_{\rm osc}$$\,=\,$3.162~au) but three different eccentricities: the proper eccentricity, $e_{\rm p}$$\,=\,$0.15, and the maximum and minimum eccentricities during the secular cycle, $e_{\rm max}$$\,=\,$0.20 and $e_{\rm min}$$\,=\,$0.10, respectively (cf. Figure \ref{fig:secular}b).
We find about a factor of 4 difference in the water sublimation rate for $e_{\rm max}$ versus $e_{\rm min}$, and about a factor of 2 difference for $e_{\rm max}$ versus $e_{\rm p}$, and for $e_{\rm p}$ versus $e_{\rm min}$, in the isothermal approximation.
The differences are not large, but could still make a significant difference for objects with activity very close to survey detection limits.
We therefore suggest that even small changes in $e_{\rm osc}$ during the secular cycle can result in significant changes in sublimation rates, and for objects exhibiting activity on the edge of detectability, even small changes in sublimation rates can have significant effects on their discoverability, i.e., making barely undetectable activity detectable and making barely detectable activity easily detectable.

Meanwhile, higher $e_{\rm osc}$ values mean that MBCs spend shorter amounts of time at higher temperatures near perihelion and longer amounts of time at lower temperatures near aphelion, affecting mantle growth rates.  Mantle growth rates depend on the duration and intensity of sublimation \citep[cf.][]{Schorghofer2008,Hsieh2015}, but it is not immediately clear whether decreases in mantle growth rates for higher-$e_{\rm osc}$ objects due to shorter times spent near perihelion are outweighed by higher peak temperatures.  If mantle growth rates are lower for higher-$e_{\rm osc}$ objects though, observable activity might be able to persist over more perihelion passages, providing more opportunities for discovery after an activation event.
Future studies to investigate such effects with detailed thermal modeling will be important to understand the consequences of dynamical evolution on the thermal evolution of MBC ice.

\subsection{Future MBC searches}

Based on the results presented here, we note that if undiscovered high-$e_{\rm osc}$ MBCs are predicted to have certain $\varpi_{\rm osc}$ values and we want to find more, we can use this information to predict when and where they might be most easily discovered, in analogy to the search for distant massive planets in our solar system \citep[e.g.,][]{Trujillo2014,Brown2016}.
Considering that we believe that currently active MBCs tend to be secularly excited in osculating eccentricity, and will therefore tend to have $\varpi_{\rm osc}$ values aligned with that of Jupiter, and are most easily discovered near perihelion (when they are most active, and also closer to the Sun and Earth, and therefore brighter), we therefore predict that new MBCs may be preferentially more likely to be discovered in the fall night sky when Jupiter's perihelion is at opposition \citep[cf.][]{Kresak1989}.

\begin{figure}
\epsscale{0.6}
\begin{center}
\plotone{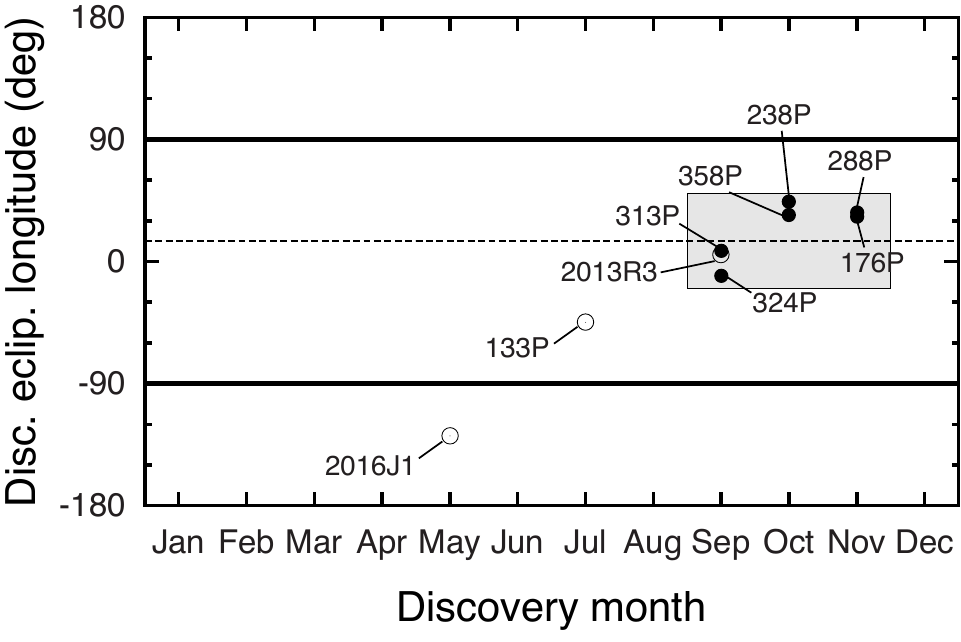}
\caption{Discovery circumstances of MBCs listed in Table \ref{tab:mbc}, in terms of heliocentric ecliptic longitude at discovery and UT month of discovery.
Galactic plane crossing occurs during June and December near $\pm90\degr$ in ecliptic longitude (thick solid lines).
Filled circles denote the sublimation-driven MBCs (S); open circles denote the MBCs whose activity may be partly rotationally-driven (S/R).
Jupiter's longitude of perihelion, $\varpi_\mathrm{J}\simeq15\degr$, is indicated by the dashed line.
The shaded region marks the concentration of the MBC discoveries (i.e. $\langle l_\mathrm{dis} \rangle \sim \varpi_\mathrm{J}$).\label{fig:discovery}}
\end{center} 
\end{figure}

To test this hypothesis, we consider the discovery circumstances of the currently known and currently active MBCs.  We find that the discoveries of MBCs that exhibit sublimation-driven activity are in fact concentrated in ecliptic longitudes near Jupiter's longitude of perihelion at a specific time of year from September to November (Figure \ref{fig:discovery}).
Application of the Rayleigh $z$ test to their ecliptic longitudes at discovery shows that the probability that these values are uniformly distributed is $<$0.001, confirming that the discoveries of currently known and currently active MBCs are clustered in ecliptic longitude at a $>$3$\sigma$ level.
This of course uses a very small sample, so this conclusion will certainly need to be revisited as more MBCs are discovered.

While search biases must ultimately be corrected for before conclusions about the distribution of main-belt ice can be drawn, we suggest that, for now, researchers could take advantage of observational selection effects to simply find more MBCs, an essential part of efforts to improve our understanding of the abundance and physical characteristics of the MBC population.
As such, we encourage observers to pay special attention to searching for new MBCs in image data obtained near opposition along the ecliptic in the fall, i.e., at ecliptic longitudes close to Jupiter's longitude of perihelion ($l_{\rm ec}$$\,\sim\,$15$\degr$).

\subsection{Inactive population}
There is of course also a large population of OMB asteroids that share similar current osculating elements as the currently active MBCs but have not been observed to be active.
Some OMB asteroids with secularly excited eccentricities may be currently active but simply not yet identified as such due to insufficiently deep observations or observations obtained at points along their orbits during which they are temporarily inactive.  Many (or most) others are likely to be genuinely inactive because, given the long-term instability against sublimation of exposed surface ice on main-belt asteroids, icy asteroids still need to experience an ``activation'' event by which surface material is removed or disrupted, excavating subsurface ice, in order for activity to be produced.  This excavation of subsurface ice could occur as the result of a collision from another asteroid, or potentially also by mass shedding caused by rotational destabilization \citep[e.g.,][]{Hsieh2009,Hirabayashi2015,Haghighipour2016}.  As localized active sites would be expected to eventually become depleted of their volatile material, such activation events would also need to have occurred relatively recently \citep[cf.][]{Hsieh2009}.
Additionally, actual composition (i.e., whether an object contains ice) certainly play a role in producing activity \citep[cf.][]{Hsieh2018}, and object size can affect the rate of activity-triggering events and production of observable activity.
\citet{Hsieh2009} pointed out that detectable active MBCs may occupy a narrow range of sizes, i.e., larger bodies present larger collisional cross-sections so would be likely to experience more frequent impacts, but also have larger escape velocities, making it more difficult for dust particles to escape, even if an activating event occurs that allows ice to sublimate.
In the case of activation by rotational instability, we expect higher activation rates for smaller bodies \citep[cf.][]{Jewitt2017}.
Future investigations considering the ratio of active to inactive objects (as determined from properly debiased survey results) will be important for constraining the rate of activating events and the typical lifetimes of active sites produced by such events.

\section{Summary}
We examine the orbital element distribution of main-belt comets (MBCs), which are objects that exhibit cometary activity yet orbit in the main asteroid belt, and may be potentially useful as tracers of ice in the inner solar system. We report the following key findings.

\begin{enumerate}

\item The currently known and currently active MBCs show an orbital clustering in longitude of perihelion, which is aligned with that of Jupiter. The clustered objects have significantly higher current osculating eccentricities than proper eccentricities, indicating that their orbits are currently (though only temporarily) secularly excited in osculating eccentricity.

\item Secular theory predicts that outer main-belt objects that currently have secularly excited osculating eccentricities tend to have current osculating angular elements which are aligned due to Jupiter's influence and this is corroborated by empirical analysis of the current osculating elements of outer main-belt asteroids.

\item At the moment, most MBCs seem to have current osculating elements that may be particularly favorable for the object becoming active (e.g., maybe because of higher perihelion temperatures or higher impact velocities causing an effective increase in the size of the potential triggering impactor population). At other times, other icy asteroids will have those favorable conditions and might become MBCs at those times as well.

\item Finally, we note that if a population of objects is predicted to have certain current osculating angular elements and we want to find more, we can use this information to predict when they might be most easily discovered.
Considering that we believe that currently active MBCs will be most easily discovered near perihelion, this means that they would be most likely found at opposition in the fall when Jupiter's perihelion is at opposition. 
The discovery circumstances of the currently known and currently active MBCs appears to support our conclusion, but it remains to be seen if future discoveries of MBCs will continue this trend.

\end{enumerate}

\acknowledgments

We thank David Jewitt, Ramon Brasser, Bojan Novakovi{\'c}, and Julio Fern{\'a}ndez for helpful comments on this manuscript.
Bojan Novakovi{\'c} provided the proper elements of MBCs.
Y.K. appreciates valuable discussions with Masateru Ishiguro and SNU colleagues.
H.H.H.\ acknowledges support from the NASA Solar System Observations program (Grant NNX16AD68G).
Y.J. is supported by funding from KASI and UNAM-DGAPA-PAPIIT.

\end{document}